\documentclass[aps,twocolumn,prl,groupedaddress,superscriptaddress,amsmath,amssymb,amsthm]{revtex4}
\usepackage{hyperref,bbm,times}
\usepackage[T1]{fontenc}
\usepackage{braket}
\usepackage{amsthm}
\usepackage{epsfig}
\usepackage{color}
\usepackage{graphicx}
\usepackage{dcolumn}
\usepackage{bm}
\usepackage[caption=false]{subfig}

\providecommand{\openone}{\leavevmode\hbox{\small1\kern-3.8pt\normalsize1}}

\usepackage{soul}

\begin{document}

\title{Enhancing nonclassical bosonic correlations in a Quantum Walk network through experimental control of disorder}

\author{Alessandro Laneve}
\email{alessandro.laneve@uniroma1.it}
\affiliation{Dipartimento di Fisica, Sapienza Universit\`{a} di Roma, Piazzale Aldo Moro, 5, I-00185 Roma, Italy}

\author{Farzam Nosrati}
\email{farzam.nosrati@unipa.it}
\affiliation{Dipartimento di Ingegneria, Universit\`{a} di Palermo, Viale delle Scienze, Edificio 9, 90128 Palermo, Italy}
\affiliation{INRS-EMT, 1650 Boulevard Lionel-Boulet,  Varennes,  Qu\'{e}bec J3X 1S2,  Canada}

\author{Andrea Geraldi}
\affiliation{Dipartimento di Fisica, Sapienza Universit\`{a} di Roma, Piazzale Aldo Moro, 5, I-00185 Roma, Italy}

\author{Kobra Mahdavipour}
\affiliation{Dipartimento di Ingegneria, Universit\`{a} di Palermo, Viale delle Scienze, Edificio 9, 90128 Palermo, Italy}
\affiliation{INRS-EMT, 1650 Boulevard Lionel-Boulet,  Varennes,  Qu\'{e}bec J3X 1S2,  Canada}

\author{Federico Pegoraro}
\affiliation{Dipartimento di Fisica, Sapienza Universit\`{a} di Roma, Piazzale Aldo Moro, 5, I-00185 Roma, Italy}

\author{Mahshid Khazaei Shadfar}
\affiliation{Dipartimento di Ingegneria, Universit\`{a} di Palermo, Viale delle Scienze, Edificio 9, 90128 Palermo, Italy}
\affiliation{INRS-EMT, 1650 Boulevard Lionel-Boulet,  Varennes,  Qu\'{e}bec J3X 1S2,  Canada}

\author{Rosario Lo Franco}
\affiliation{Dipartimento di Ingegneria, Universit\`{a} di Palermo, Viale delle Scienze, Edificio 6, 90128 Palermo, Italy}

\author{Paolo Mataloni}
\affiliation{Dipartimento di Fisica, Sapienza Universit\`{a} di Roma, Piazzale Aldo Moro, 5, I-00185 Roma, Italy}

\begin{abstract}
The presence of disorder and inhomogeneities in quantum networks has often been unexpectedly beneficial for both quantum and classical resources. Here, we experimentally realize a  controllable inhomogenous Quantum Walk dynamics, which can be exploited to investigate the effect of coherent disorder on the quantum correlations between two indistinguishable photons. Through the imposition of suitable disorder configurations, we observe two photon states which exhibit an enhancement in the quantum correlations between two modes of the network, compared to the case of an ordered Quantum Walk. Different configurations of disorder can steer the system towards different realizations of such an enhancement, thus allowing spatial and temporal manipulation of quantum correlations.
\end{abstract}

\date{\today }

\maketitle

\textit{Introduction.---} A thorough characterization of genuine quantum traits is crucial to understand the boundary between classical and quantum phenomena \cite{zurek2003decoherence}, and to perform quantum information tasks \cite{nielsen2002quantum}. To this aim, several quantification methods have been introduced to faithfully identify the presence of quantum (nonclassical) resources such as entanglement \cite{horodecki2009quantum}, coherence \cite{streltsov2017colloquium}, discord \cite{ModiRMP}, joint measurability \cite{quintino2014joint}, steering \cite{wiseman2007steering}, or thermal operations \cite{ng2018resource} in the case of composite systems.  

Indistinguishability of quantum identical particles \cite{wiseman2003entanglement,franco2018indistinguishability} has also revealed as a useful nonclassical resource. From an operational point of view, particles are so-called indistinguishable if they are in the same mode with respect to a characterization via two-particle interference \cite{ou1987relation}. From a broader perspective, the indistinguishability concept is related to a given set of quantum measurements \cite{nosrati2020robust}. In fact, indistinguishability plays a fundamental role in raising quantum processes, such as many-body interference \cite{ou1987relation, giordani2018experimental}, entanglement generation \cite{wiseman2003entanglement, franco2018indistinguishability, benatti2020entanglement,bose2002generic,barros2020entangling,sun2020experimental}, quantum teleportation \cite{sun2020experimental}, quantum metrology \cite{dowling2008quantum, giovannetti2011advances}, quantum coherence \cite{mandel1991coherence,sperling2017quantum, castellini2019indistinguishability}, quantumness protection \cite{perez2018endurance, nosrati2020robust, nosrati2020dynamics}, quantum key distribution \cite{lo2012measurement,liu2013experimental}, and the high state complexity exploited by Boson Sampling algorithms \cite{aaronson2011computational,zhong2020quantum}.  

In this context, it is important to understand how quantum features based on indistinguishability behave in a dynamical framework, specifically in the case of bosons propagating through a non-homogeneous system.
 For a large variety of systems, the disorder plays a detrimental role because it drives the system into decoherence \cite{kendon2007decoherence}. Contrarily, for some systems, the disorder can enhance physical properties such as coherent transport \cite{mohseni2008environment},  quantum algorithms speedup \cite{kendon2003decoherence}, and quantum correlations \cite{vieira2013dynamically, vieira2014entangling, zeng2017discrete, wang2018dynamic}. These effects commonly appear due to the interaction with an external environment, though not always featuring a back-action mechanism \cite{franco2012revival, xu2013experimental}. \\
A suitable theoretical platform to perform such a study is represented by Quantum Walk (QW), which provides a very general coherent propagation model: at variance with classical Random Walks, QWs are able to preserve genuine nonclassical features such as superposition, interference, and entanglement \cite{aharonov1993quantum,ambainis2001one,venegas2012quantum}. QWs  provide powerful models to describe energy transport phenomena in different types of systems like photosynthetic complexes  \cite{mohseni2008environment, plenio2008dephasing}, or solid state ones, as in the case of Luttinger Liquids \cite{bulchandani2020superdiffusive}.

It has been shown that adjustable disorder plays a significant role in the evolution of quantum walkers in which the ballistic growth can become anomalous, classical, or localized \cite{schreiber2011decoherence, crespi2013anderson, de2014quantum, geraldi2019experimental, geraldi2020subdiffusion}. The dynamics of a quantum walker is intimately connected to its nonclassical features. The way quantum-correlated walkers, realized by photon pairs, evolve in a homogeneous optical lattice has been investigated, highlighting the different behavior of distinguishable or indistinguishable photons \cite{bromberg2009quantum, peruzzo2010quantum, lahini2012quantum, sansoni2012two, poulios2014quantum}. Moreover, the spreading pattern of the quantum walker(s) can be modified through various types of disorder \cite{lahini2010quantum, crespi2013anderson, geraldi2019experimental,geraldi2020subdiffusion}.

\begin{figure}[t!]
	\includegraphics[width=\columnwidth]{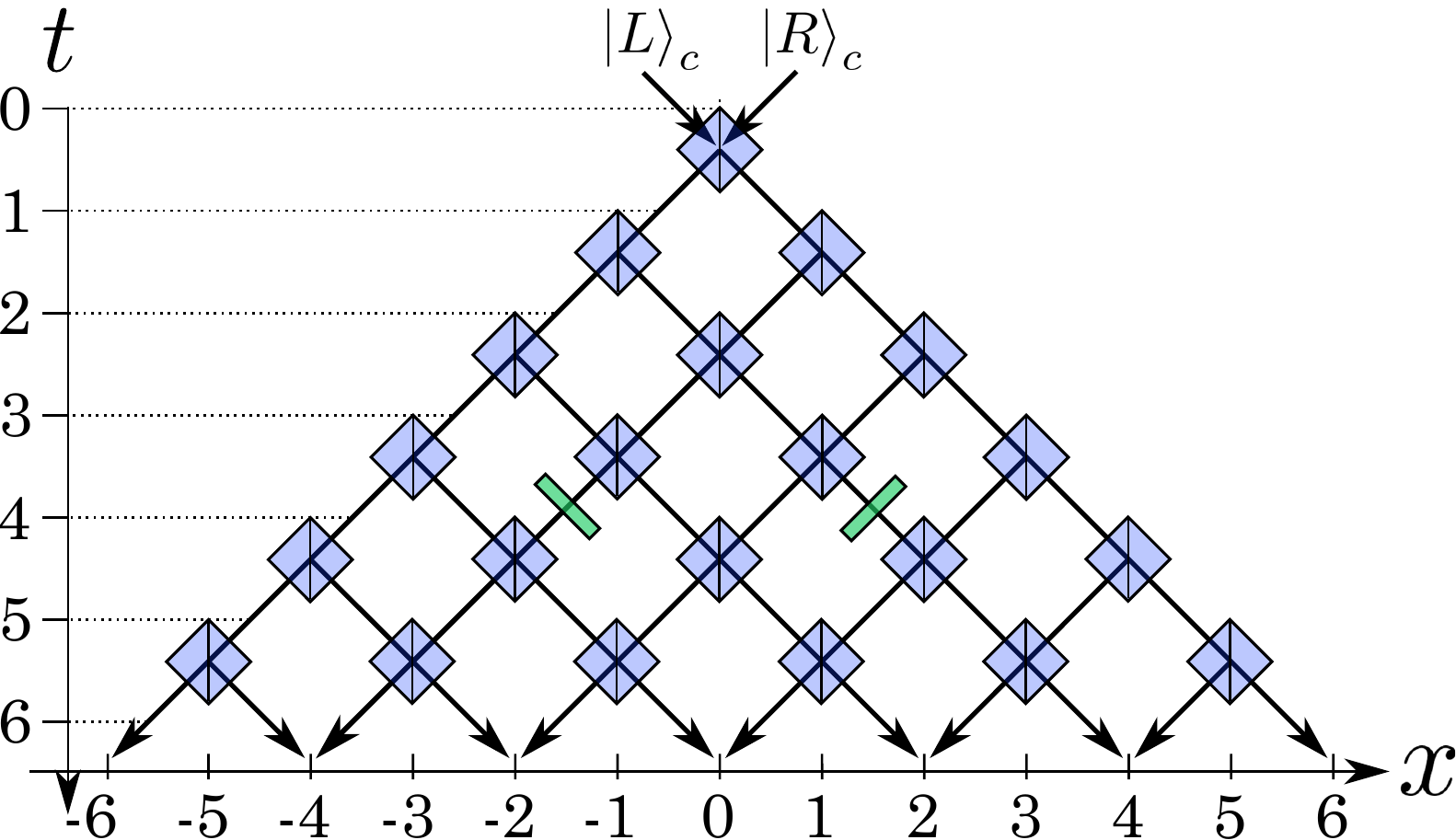}
	\caption{\textbf{Network representation of a disordered QW according to $p$-diluted model.} The green bars stand for the presence of $\pi$ phases on the  path which can be added to or removed from the network. Coin and shift operations are here represented by Beam Splitters (shaded cyan squares), just as in the actual experimental realization. The $\ket{R}_c$ and $\ket{L}_c$ states are indicated.}\label{fig:network}
\end{figure}

\begin{figure*}[t] 
\centering
\includegraphics[width=0.85\textwidth, height=0.32\textwidth]{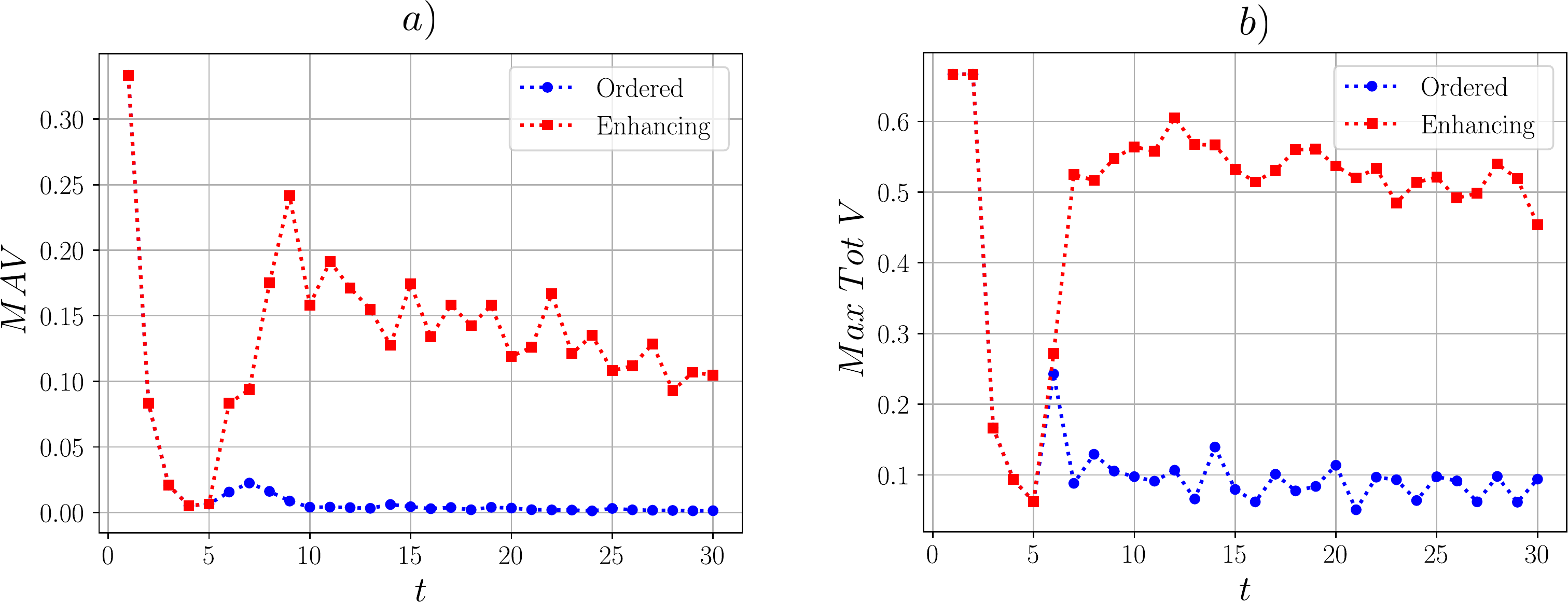}
\caption{\textbf{Numerical simulation compression between ordered and enhancing disordered QW}. \textbf{a)} Maximum Achievable Violation (MAV) and \textbf{b)} maximum achievable Total Violation versus the number of steps (discrete time) $t$  for the ordered (blue circles)  and enhancing disordered (red squares) QW.} \label{fig:viol_teo_trends}
\end{figure*}

At a variance with previous studies so far \cite{wang2018dynamic}, it still remains to find strategies for enriching two-particle quantum correlations via disorder control. Here, we fill this gap through the experimental observation of the propagation of two indistinguishable photons (biphoton) in a one-dimensional inhomogeneous Discrete Time Quantum Walk (1D DTQW). We demonstrate that, by merely imposing some specific disorder configurations while keeping the system isolated, it is possible to significantly retrieve the initial non-classicality of the system, after a certain number of evolution steps. On the other hand, we numerically show that, on average, the imposition of random disorder configurations results in a decrease of the initial biphoton quantum correlations (details are reported in the Supplementary Material (SM)).  Our findings unambiguously prove that the presence of disorder fosters the dynamical enhancement of biphoton quantum correlations in a controllable fashion, paving the way to its potential employment in quantum information scenarios.

\textit{Theoretical framework.---} The 1D QW model consists of one or more walkers coherently moving along the discretized sites of a line \cite{ambainis2001one}.  In general, the state of the system can always be written as a superposition of the QW modes $\ket{\Psi(t)}=\sum_k \alpha_k(t) \ket{k}$, where each mode $\ket{k}:=\ket{x} \ket{\sigma}$ is defined by both position $\ket{x}$ and its coin $\ket{\sigma}=\{\ket{L},\ket{R}\}$, and amplitudes $\alpha_k(t)$  depend on the past evolution of the walker. Therefore, the single step evolution can be written as
\begin{equation}
       \ket{\Psi(t+1)}=\sum_k e^{i\phi_k(t)} \hat{U}\alpha_k(t)\ket{k},
\end{equation}
where $\hat{U}=\hat{S}\cdot(\hat{I}\otimes\hat{C})$ is the one-step evolution operator with $\hat{S} = \sum_x \ket{x+1}\bra{x}\otimes\ket{L}\bra{L}+\ket{x-1}\bra{x}\otimes\ket{R}\bra{R}$ being the shift operator, that  moves the walker according to the coin state, $\hat{I}$ the position identity operator and $\hat{\mathcal{C}}$ the coin operator, which in our case reads $\hat{\mathcal{C}}=\frac{1}{\sqrt{2}}\left(\ket{L}\bra{L}+i\ket{L}\bra{R}+i\ket{R}\bra{L}+\ket{R}\bra{R}\right)$. Here, step-position dependent phases $\phi_k(t)$, out of two choices $0$ or $\pi$, are responsible for the dynamical disorder that the quantum walker experiences, according to the so-called $p$-diluted model \cite{geraldi2019experimental}. Now, the question arises how the application of the phase shiftings $\pi$, represented by green bars in Fig.~\ref{fig:network}, in different positions or at different steps influences the way quantum correlations propagate in the network structure underlying the QW evolution (clearly represented by Fig.~\ref{fig:network}).\\
In order to study the effect of disorder over nonclassical bosonic correlations in a QW dynamics, we consider two indistinguishable photon walkers as input. This choice is strategical since, unlike states of distinguishable photons, the state of an indistinguishable photon pair (biphoton) $\ket{\Psi^{(2)}}=\ket{k_1, k_2}$ ($\ket{k_i} = \ket{x_i}\ket{\sigma_i}$) exhibits intrinsic quantum correlations \cite{franco2016quantum,compagno2018dealing}. The two-particle evolution is obtained by applying $\hat{U}\otimes \hat{U}$ to the initial state. Inspired by classical intensity correlation of light, one way to measure the nonclassicality of the correlation between two detected outputs is by violating the Cauchy-Schwarz inequality \cite{bromberg2009quantum, peruzzo2010quantum}
\begin{equation}
    V_{ij}=\frac{2}{3}\sqrt{\Gamma_{ii}\Gamma_{jj}}-\Gamma_{ij}< 0,
    \label{eq:violation}
\end{equation}
where $\Gamma_{ij}$ is the probability of finding a photon in mode $i$ and the other one in $j$, namely the probability of measuring a coincidence between modes $i$ and $j$. Inequality \eqref{eq:violation}, in fact, stands for classically correlated light and its violation is assumed to witness and quantify the presence of quantum correlations, as a signature of photon bunching \cite{bromberg2009quantum, peruzzo2010quantum, poulios2014quantum}.

\textit{Numerical Results.---} Preliminary simulations were carried on in an ideal $p$-diluted  framework: two indistinguishable photons travelling a bulk-optics 1D DTQW, provided with space-time disorder. As displayed in detail in SM, the correlations of the system show a decrease in non-classicality on average, as disorder increases. On the other hand, it is possible to find specific disorder configurations which produce an enhancement in non-classicality. In order to find such phase maps, we simulate the evolution with $10^4$ different phase maps for each step number $t$ up to 30 steps, for a total amount $3\cdot10^5$  of explored configurations. We compute the $V_{ij}$ between each pair of output modes for each simulated probability distribution, obtaining the corresponding violation matrices. Then, the entire set of $V_{ij}$ elements corresponding to any combination of $i$ and $j$ modes are reported. Now, we can compare the values for any pair $(i,j)$ and each phase map at a given step, to find the Maximum Achievable Violation (MAV), i.e. the maximum positive value of $V_{ij}$ which could be achieved at that given step (we name this as optimal two-mode quantum correlation). Simulation results are shown in Fig.~\ref{fig:viol_teo_trends} \textbf{a)} and \textbf{b)}, respectively. We report the MAV as a function of the number of steps, and the step-wise trend of the maximum achievable Total Violation, defined as the sum of all the positive values $V_{ij}$  of the considered violation matrix  (named here as total quantum correlation).  
Since after the 7th step the explored configurations do not cover the entire set of possible disorder patterns, the results can not be considered absolutely optimal, but rather enhancing in comparison with the ordered case.
Nevertheless, the analysis highlights that disorder helps to retrieve quantum correlations after a specific step of QW. Hence, the MAV starts growing compared to the standard ordered case, reaching a peak at the $9$th step in our framework. Similar, but not identical, results were obtained for the system's maximum Total Violation, suggesting that the two quantities are related but not bound to be maximized together. From numerical results, we can conclude that disorder, acting through mere interference, significantly modifies the evolution of the walker, not only reshaping the probability distribution but also affecting the amount of quantum correlation between the photons. As a consequence, disorder may enable enhancement of the quantum correlation of a bipartite system. It is worth noting that no violations are observed, whatever the phase map, when a state of distinguishable photons is employed \cite{peruzzo2010quantum}.

\begin{figure}[!t]
 	    \centering
 	    \includegraphics[width=1\columnwidth]{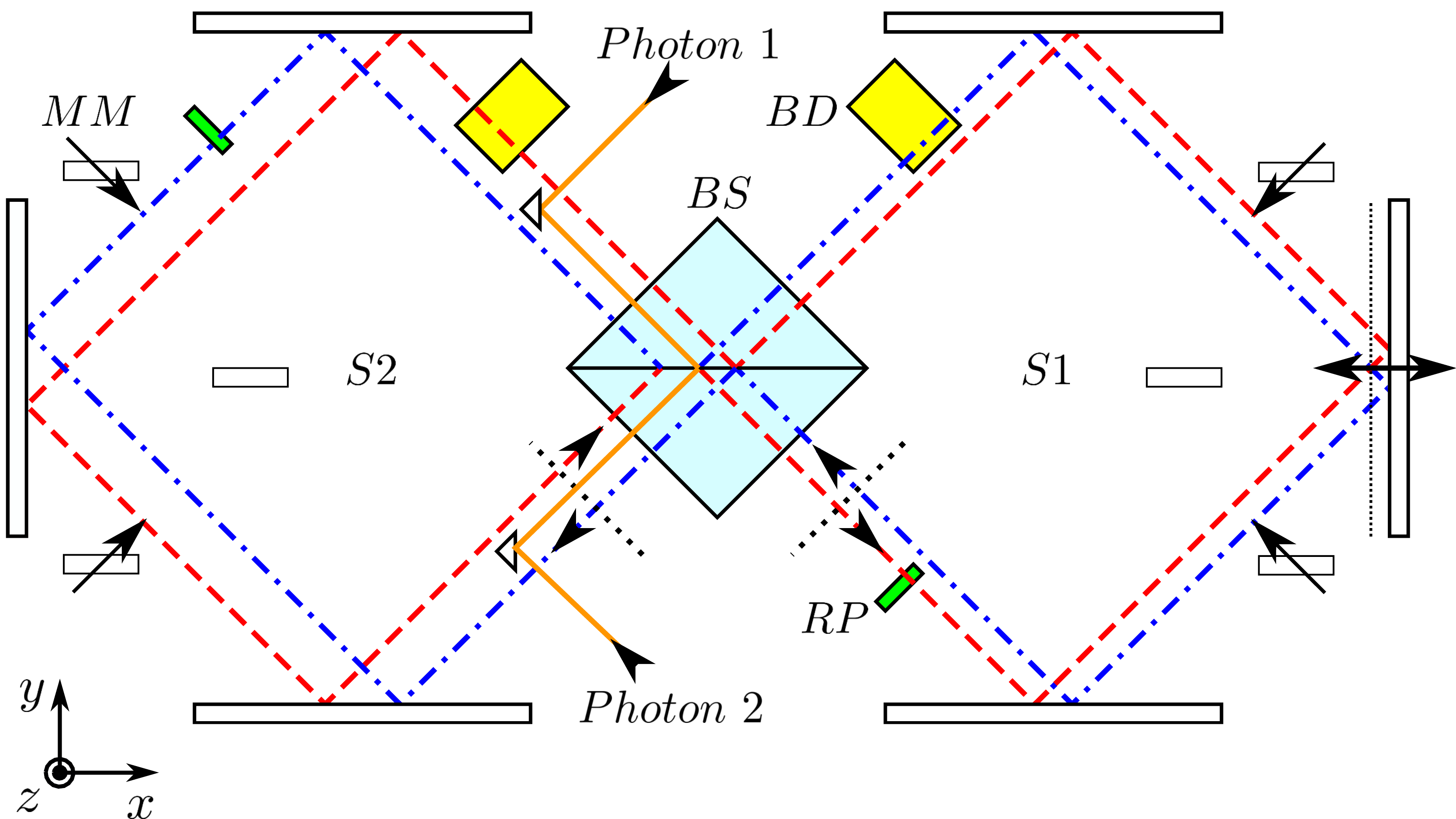}%
 	    \caption{\textbf{2D sketch of the experimental setup.} BS: beam splitter, BD: beam displacer, RP: rotating glass plates, MM: moving mirror, S1: Sagnac Interferometer for odd steps, S2: Sagnac Interferometer for even steps. Blue and red beams circulate in opposite directions and impinge on the BS in the same horizontal point but at different heights along the z direction, due to the effect of BDs.}
 	    \label{fig:setup}
\end{figure}

\textit{Experimental Setup.---} The experimental setup, designed to test the theoretical predictions, consists of a bulk-optics multipass double Sagnac Interferometer (SI), already exploited in a previous experiment \cite{geraldi2019experimental}, in which inhomogeneities, described by the phases $\phi_k(t)$, can be addressed independently both in step number and evolution mode. The bulk-optics setup shown in Fig.~\ref{fig:setup} is analogous to a chain of intrinsically phase-stable Mach-Zehnder Interferometers (MZIs), each of them provided with an individually tunable phase shifting. The additional exploitation of the $z$ direction allows to effectively realize a Beam Splitter (BS) network, such as the one in Fig.~\ref{fig:network}, which reproduces a 1D DTQW dynamics where the coin and position states are both encoded in the input and output direction of photons with respect to the BS. Thanks to the particular geometry of the implementation, each propagation mode of the QW at each given step has a specific position in the plane transverse to the propagation direction. Therefore, the phase shifts can be independently addressed in each mesh of the QW by the simple insertion of Rotating Glass Plates (RP) along the propagation path. The output state can be measured through a set of Moving Mirrors (MMs), intercepting and extracting from the setup only modes of the selected step $t_j$. Also, the previous propagation steps $t<t_j$ are not affected in any way by the measurement procedure.  The extracted radiation is then coupled to a single-mode fiber and measured (for further details on the setup see Ref. \cite{geraldiQW2019,geraldi2019experimental}). Using couplers to collect the extracted photons, we can measure coincidences between all possible modes at each step and experimentally reconstruct the two-photon probability distribution.

\begin{figure*}[!t] 
\centering
\includegraphics[width=0.85\textwidth]{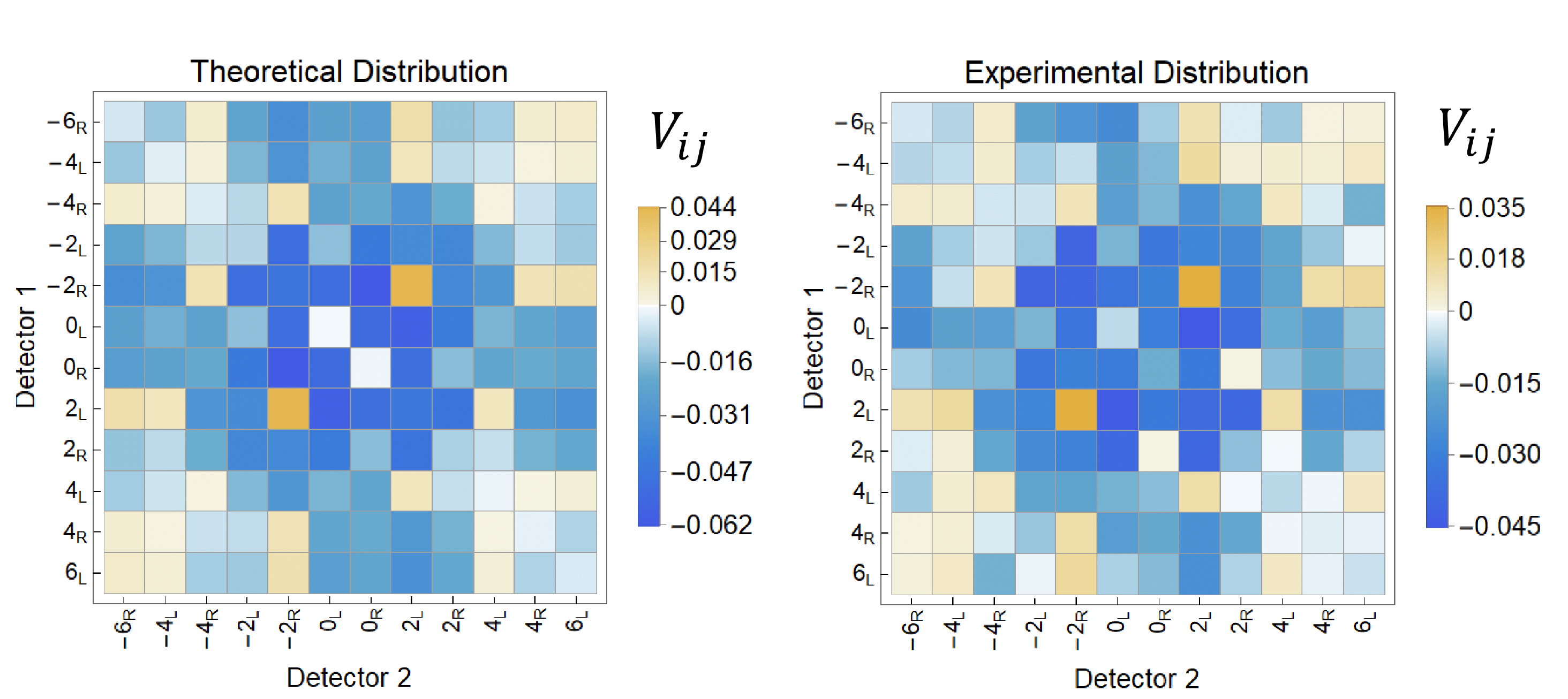}
\caption{\textbf{Comparison between the theoretical and experimental violation matrices at the 6th step for an enhancing disorder configuration}. Numerical simulations are performed taking to account experimental parameters. The expected peak in the value of $V$ is experimentally found, while the measured output coincidence distributions reach globally a similarity value of $97.5(\pm1.3)\%$. } \label{fig:6step_exp_violmat}
\end{figure*}

\begin{figure*}[t] 
\centering
\includegraphics[width=0.85\textwidth, height=0.32\textwidth]{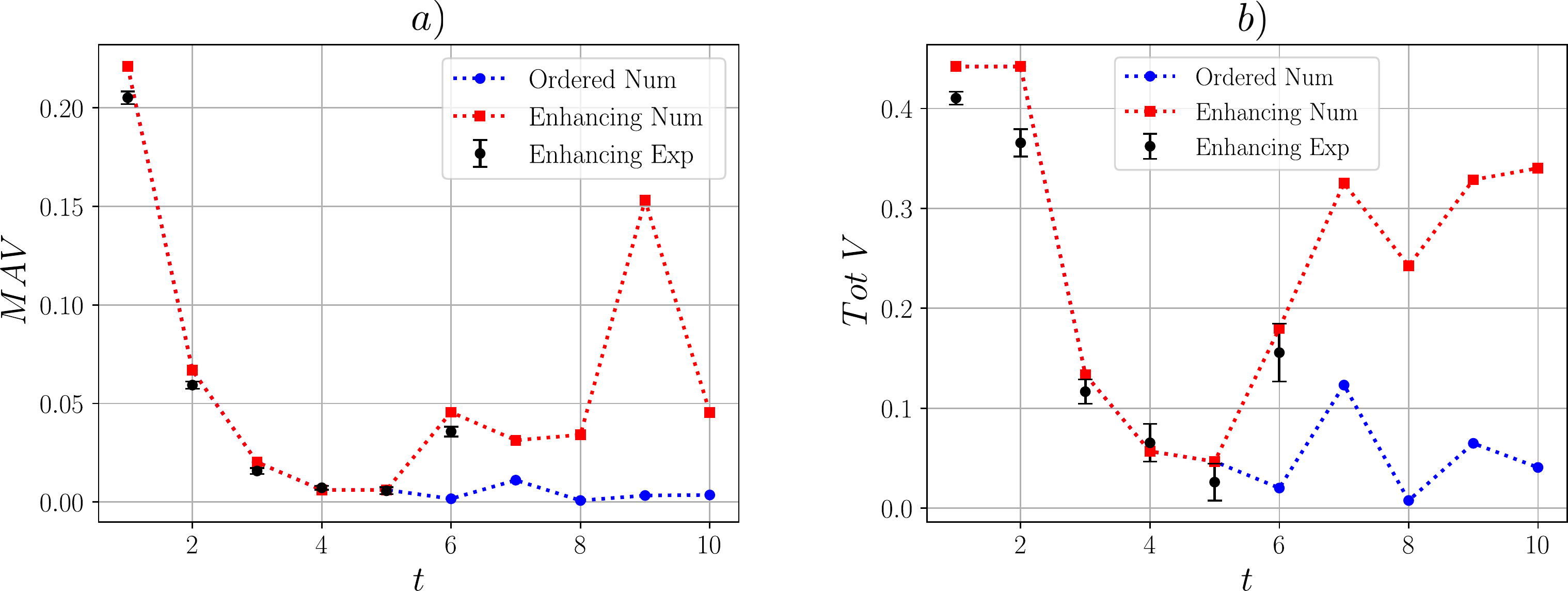}

\caption{\textbf{Experimental results compression between the order and the enhancing disordered configuration QW.} Experimental results for (dark dot) \textbf{a)} Maximum Achievable Violation (MAV) and \textbf{b)} corresponding Total Violation versus the number of steps. The trends are compared with the simulation for enhancing disorder (red squares) and ordered evolution (blue circles). The expected results are obtained by numerical simulations performed accounting for experimental parameters, so that the theoretical trends show some discrepancies with respect to the ones in Fig.~\ref{fig:viol_teo_trends}.}\label{fig:viol_comparison}

\end{figure*}

By post-selecting on the coincidences, there is no need for heralding procedures, so losses are crucially reduced, making it possible to measure the two-particle distribution while keeping a high tunability of network parameters, even up to the $6$th steps.

\textit{Experimental Results---} To experimentally verify disorder-induced changes in the violation matrix, we measure both ordered and disordered evolution QW output distributions. In fact, based on the simulation study displayed in Fig.~\ref{fig:viol_teo_trends}, the first quantum correlation enhancement, due to disorder, shall occur at the $6$th step of QW. Therefore, the output violation matrices for optimal phase maps are measured up to the $6$th step. Since there is no enhancement until the 5th step, the corresponding optimal phase maps can be considered equivalent to the ordered one, while for the 6th step it is possible to find many specific disorder configurations enhancing the non-classicality in the correlation between two chosen output modes, both for the ideal case and accounting for experimental parameters.
The phase map selected for the experimental implementation features phase shifters at step $t=4$, position $x=-2$ with coin $\sigma=L$ and $x=2$ with coin $\sigma=R$ set to $\pi$, while all the others are left to zero; the corresponding experimental output violation matrix is shown in Fig.~\ref{fig:6step_exp_violmat}, compared with the expected one, where the mode $\ket{k}=\ket{x} \ket{\sigma}$ is indicated by $x_\sigma$. A strong quantum correlation peak appears at modes $\left(2_L,-2_R\right)$ and $\left(-2_R,-2_L\right)$ confirming the expectation.
As a further relevant result, the experimental step-wise trend for MAV is shown in Fig.~\ref{fig:viol_comparison}$a$, in comparison with the expected enhanced one obtained by numerical analysis, taking into account experimental constraints. They are plotted together with the ordered case trend to provide a clear display of the beneficial effect of the non-homogeneous evolution. Theoretical patterns are shown up to the $10$th step, as a reference. The corresponding trends for the Total Violation computed over the same output distributions are also reported in Fig.~\ref{fig:viol_comparison}$b$. Simulations of the MAV values in the ordered case show that the quantum correlation spreads in a homogeneous network, so that the values of $V_{ij}$ and the Total Violation are going to decline as the propagation proceeds. However, as can be seen in Fig.~\ref{fig:viol_comparison}$a$, the inhomogeneity enriches the quantum correlation between two indistinguishable photons at the given modes.  Experimental evidences, reported in Fig.~\ref{fig:viol_comparison}$a$, show that the very same configuration also enhances the total quantum correlation of the quantum walk. Discrepancies between theoretical trends shown in Fig.~\ref{fig:viol_comparison} and in Fig.~\ref{fig:viol_teo_trends} are due to asymmetries in the experimental setup, specifically the exploitation of an unbalanced BS ($R=45/T=55$). Experimental errors are derived from the Poissonian statistics of the measured coincidences. Deviations from the expected results are mainly due to modest drops in photons indistinguishability along the evolution, caused by small alignment imperfections, which become more significant as the travelled free space increases. Nevertheless, the indistinguishability decline slightly affects the exact violation values, while not changing the overall trend.

\textit{Conclusion.---} The presented numerical and experimental analysis demonstrates that two-mode quantum correlations due to particle indistinguishability, which disperse through the lattice and rapidly decay in an ordered evolution, can be retrieved after a minimum evolution time by inserting suitable inhomogeneity patterns in the system. By changing the disorder configurations, it is possible to tune the two-mode and total enhancement of non-classicality in position and intensity; this corresponds to an adaptive network whose parameters evaluation determines the focusing of nonclassical resources in selected modes. Also, we show that the two-mode quantum correlation diminishes in the case of random phase disorder in the system (for details see SM). Nevertheless, this quantum correlation degradation can be challenged by single realizations of disorder. These results supply a conceptual and practical advance compared to previous studies limited to single-photon disorder-assisted quantum correlation enhancement between two degrees of freedom of the photon \cite{wang2018dynamic}. In fact, since violations of Eq.~\ref{eq:violation} indicate biphoton quantum correlations between two modes, our method can be especially promising for Quantum Metrology issues.
It is yet to understand whether this enhancement procedure can be generalized to systems with $N>2$ photons or not, which could result in a benchmarking outcome in the context of Quantum Resource Theories.

\textit{Acknowledgments.---} We acknowledge support from the European Commission Grants No. FP7-ICT-2011-9-600838 (QWAD–Quantum Waveguides Application and Development). We thank Fabio Sciarrino for useful help and discussions.

\bibliography{ref}
\newpage

\clearpage

\section{Supplemental Material}

\section{$p$-Diluted Network} 
Here, we briefly review the $p$-Diluted network model of quantum walk dynamics \cite{geraldi2019experimental}. As explained in the main text, the quantum state of the system can be written as a superposition of the QW modes $\ket{\Psi(t)}=\sum_k \alpha_k(t) \ket{k}$, where each mode $\ket{k}:=\ket{x} \ket{\sigma}$ is defined by both position $\ket{x}$ and its coin $\ket{\sigma}=\{\ket{L},\ket{R}\}$, and amplitudes $\alpha_k(t)$  depend on the past evolution of the walker. Therefore, the single step evolution can be written as
\begin{equation}
       \ket{\Psi(t+1)}=\sum_k e^{i\phi_k(t)} \hat{U}\alpha_k(t)\ket{k},
\end{equation}
where $\hat{U}=\hat{S}\cdot(\hat{I}\otimes\hat{C})$ is the one-step evolution operator  with $\hat{S} = \sum_x \ket{x+1}\bra{x}\otimes\ket{L}\bra{L}+\ket{x-1}\bra{x}\otimes\ket{R}\bra{R}$ being the shift operator, that  moves the walker according to the coin state and $\hat{\mathcal{C}}=\frac{1}{\sqrt{2}}\left(\ket{L}\bra{L}+i\ket{L}\bra{R}+i\ket{R}\bra{L}+\ket{R}\bra{R}\right)$ the coin operator. Step-position dependent phases $\phi_k(t)$ are chosen out of two choices $0$ or $\pi$. In the main text, we focused on the single \textit{phase map} that enhances the quantum correlation between the walkers. Here, we assume that the phases experienced by the quantum walker during its evolution are independently and randomly generated. In order to generate the random \textit{phase map}, a Bernoulli process determines whether the phase shift at step $t$ and position $x$ is  $\phi=0$ (with a certain probability $(1-\frac{p}{2})$) or $\phi=\pi$ (with probability $\frac{p}{2}$). In the end, a phase map is a set of matrices $\{\Phi_p(t)\}_{t=0,...t_{max}}$ which describes all the phases imposed on the walker during the QW process with a given fixed $p$. On average, photons will experience a $\frac{p}{2}$ percentage of "flipped" phases during the quantum evolution. It is worth noting that the evolution stays coherent: in fact, there's no decoherence in this framework; if the initial state is pure, the final state stays so.

\begin{figure*}[!t] 
\centering
\includegraphics[width=0.95\textwidth]{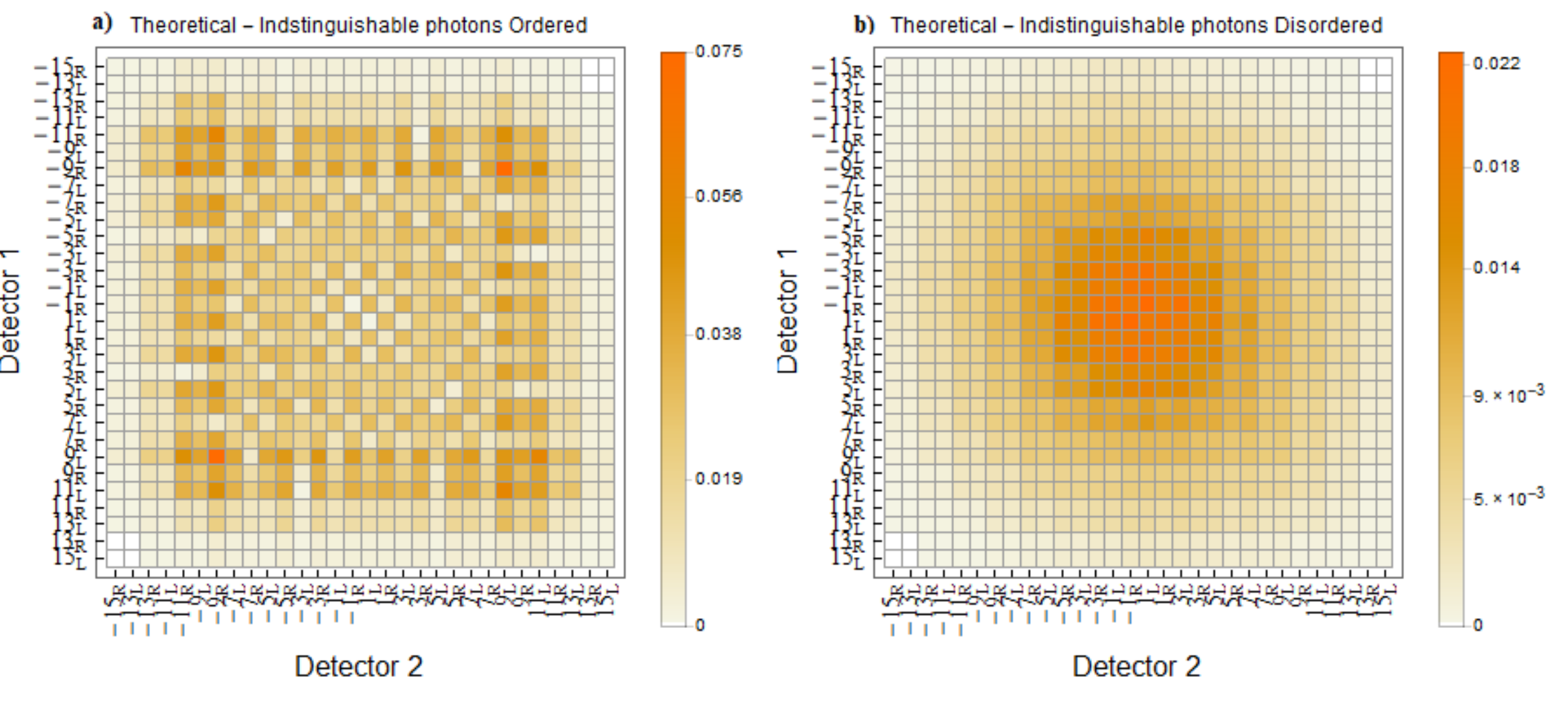}
\caption{\textbf{Numerical simulation of coincidences matrices}. Step 15 output distributions of indistinguishable photons in the \textbf{a)} ordered and completely \textbf{b)} disordered case. The disordered matrix has been computed by averaging over $10000$ disorder configurations.} \label{fig:ind_distr}

\end{figure*}

\begin{figure*}[!t]
\centering
\includegraphics[width=1.00\textwidth]{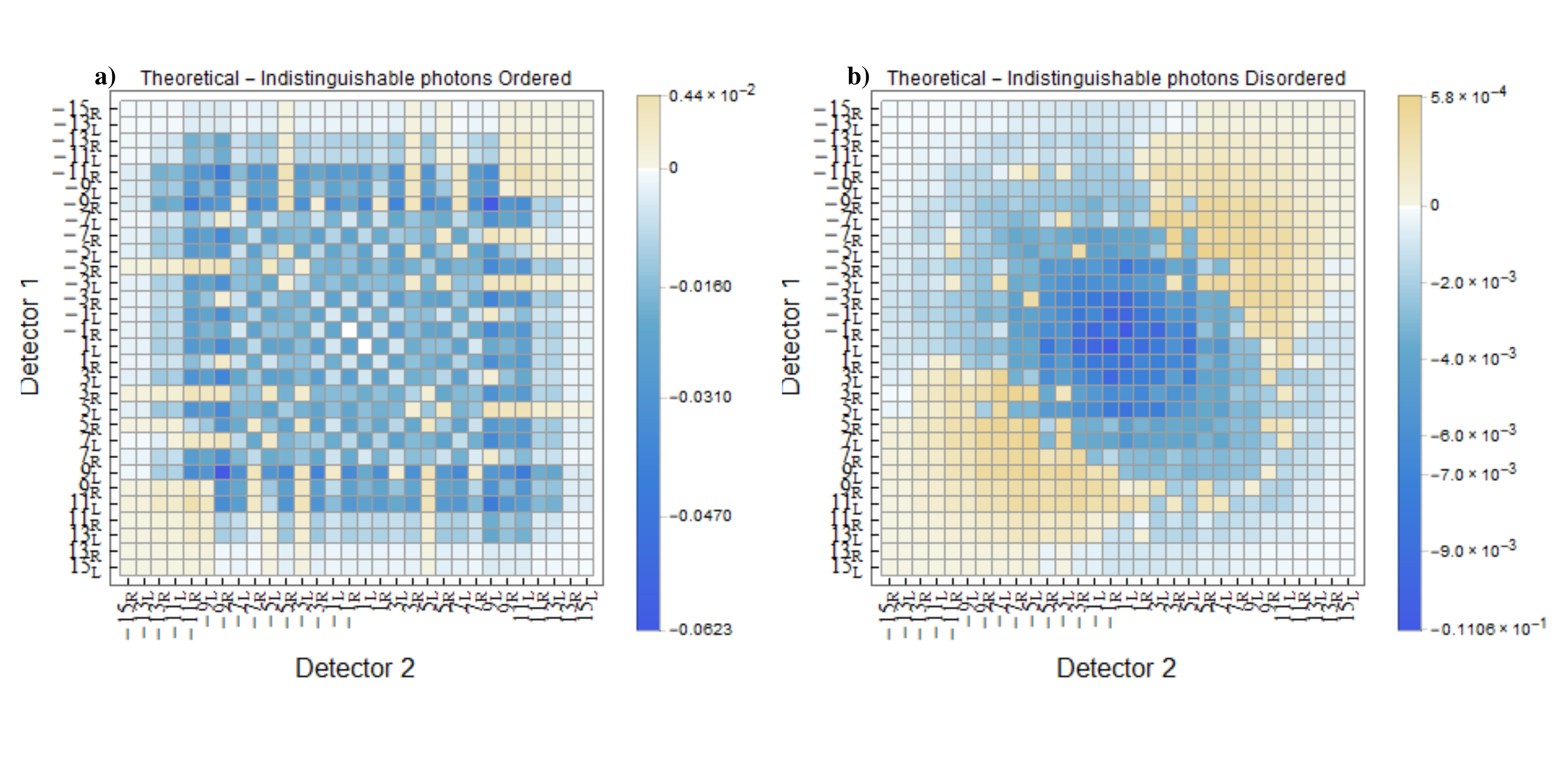}
\caption{\textbf{Numerical simulation of violation matrices}. Step 15 matrices of the $V_ij$ values of indistinguishable photons in the \textbf{a)} ordered and completely \textbf{b)} disordered case. The disordered matrix has been computed by averaging over $10000$ disorder configurations.}
\label{fig:ind_viol}

\end{figure*}

%

\section{Quantitative analysis of violation}

The aim of this section is to provide a clearer interpretation of the quantity $V_{ij}$, which, in the main text, is exploited as a quantifier for nonclassicality in bosonic correlations. In particular, we focus on the relationship lying between this quantity and the two-particle boson bunching, the most simple and straightforward effect of bosonic indistinguishability; then, this relationship is extended to the network case, fitting the experimental implementation presented in the main text.\\
As already described in the main text, the violation of the inequality:
\begin{equation}
    \frac{2}{3}\sqrt{\Gamma_{i,i}\Gamma_{j,j}}-\Gamma_{i,j}<0
\label{eq:viol}
\end{equation}
allows to point out the presence of non-classical correlations in photonic systems \cite{bromberg2009quantum,lahini2010quantum,lahini2012quantum,poulios2014quantum}.
The meaning of this simple relation can be traced back to the result of an elemental boson bunching phenomenon, i.e. the HOM effect \cite{hong1987measurement}, for the case of non perfectly indistinguishable photons.
We consider a photon pair with a given probability $q$ of being distinguishable, travelling through a supposedly balanced $BS$. After the $BS$, photons can be measured in three possible combinations of the two output modes and the probability of each combination depends on the probability of the photons being indistinguishable (which is $1-q$). Therefore, inequality \eqref{eq:viol} can be written as:
\begin{equation}
    \frac{2}{3}\bigg(\frac{1}{2}-\frac{q}{4}\bigg)-\frac{q}{2}<0
\label{eq:viol_prob}
\end{equation}
which, in order to be violated, requires a value $q<\frac{1}{2}$, corresponding to photons which are more likely to be indistinguishable than distinguishable.
Hence, the inequality \ref{eq:viol} provides a straightforward quantifier of the effective indistinguishability of photons, in the operational context of boson bunching occurrence.
In the general case of a $BS$ network, such as the one exploited in the present work, a value of $V_{ij}>0$ can be subject to multiple interpretations.
In the general case for a pure initial state, after a $n$ step propagation the system will be in a superposition state which can be written considering the number of photons travelling in the modes of interest:
\begin{align*}
    \ket{\Phi}=\sqrt{1-\Pi}\big(...\big)+\sqrt{\Pi}\big(\alpha_{k_1 k_2} \ket{2}_{k_1}\ket{0}_{k_2}+\\
    +\alpha_{k_1 k_2} \ket{1}_{k_1}\ket{1}_{k_2}+\alpha_{k_2 k_2} \ket{0}_{k_1}\ket{2}_{k_2}\big)
\end{align*}
where $\Pi$ is the overall probability of having both photons in the selected modes, which normalizes the $\alpha_{ij}$ coefficients, while $\{\ket{k_1},\ket{k_2}\}$ are the two output modes under observation, corresponding to combined states of position and coin of the form $\ket{k}:=\ket{x} \ket{\sigma}$. It is possible, obviously, also to have single photon states of the two modes, but they would be invisible to coincidence-like measurements.
In this case, the amount of violation between modes $k_1$ and $k_2$ can be computed as:
\begin{equation}
    V_{k_1 k_2}=\Pi*\bigg(\frac{2}{3}\sqrt{|\alpha_{k_1 k_1}|^2|\alpha_{k_2 k_2}|^2}-|\alpha_{k_1 k_2}|^2\bigg)
\label{eq:violation}
\end{equation}
Therefore, the violation amount depends on two factors:
\begin{itemize}
    \item the actual non-classicality of the correlation determining a positive or negative value
    \item the global probability of the selected output modes (given by $\Pi$)
\end{itemize}
The first factor is the one pointing out the form of a hypothetically post-selected state of the two photons emerging from the considered modes. The higher this factor, the cleaner is the distillation of NOON states by post-selection, since it necessarily corresponds to a low $|\alpha_{k_1 k_2}|^2$.
The second factor is an amplification parameter, which gives the probability of actually finding two photons in the two-modes selected subsystem, hence it gives the efficiency of the NOON states distillation.
In conclusion, the violation value provides an indication over the composite effect of the two parameters, hence its maximization can be related to either one or the other. Hence, this aspect needs to be taken into account in a hypothetical application of this protocols.
For instance, the most external output modes will provide the most pure NOON states, since they are the mere propagation of the first HOM resulting state, but with a very low probability. On the other hand, by means of disorder, it is possible to manipulate the probability for central modes and get an higher efficiency, but at the cost of a non-zero chance of extracting a useless state.\\

\section{Numerical results}
\begin{figure}[!t] 
\centering
\includegraphics[width=0.45\textwidth]{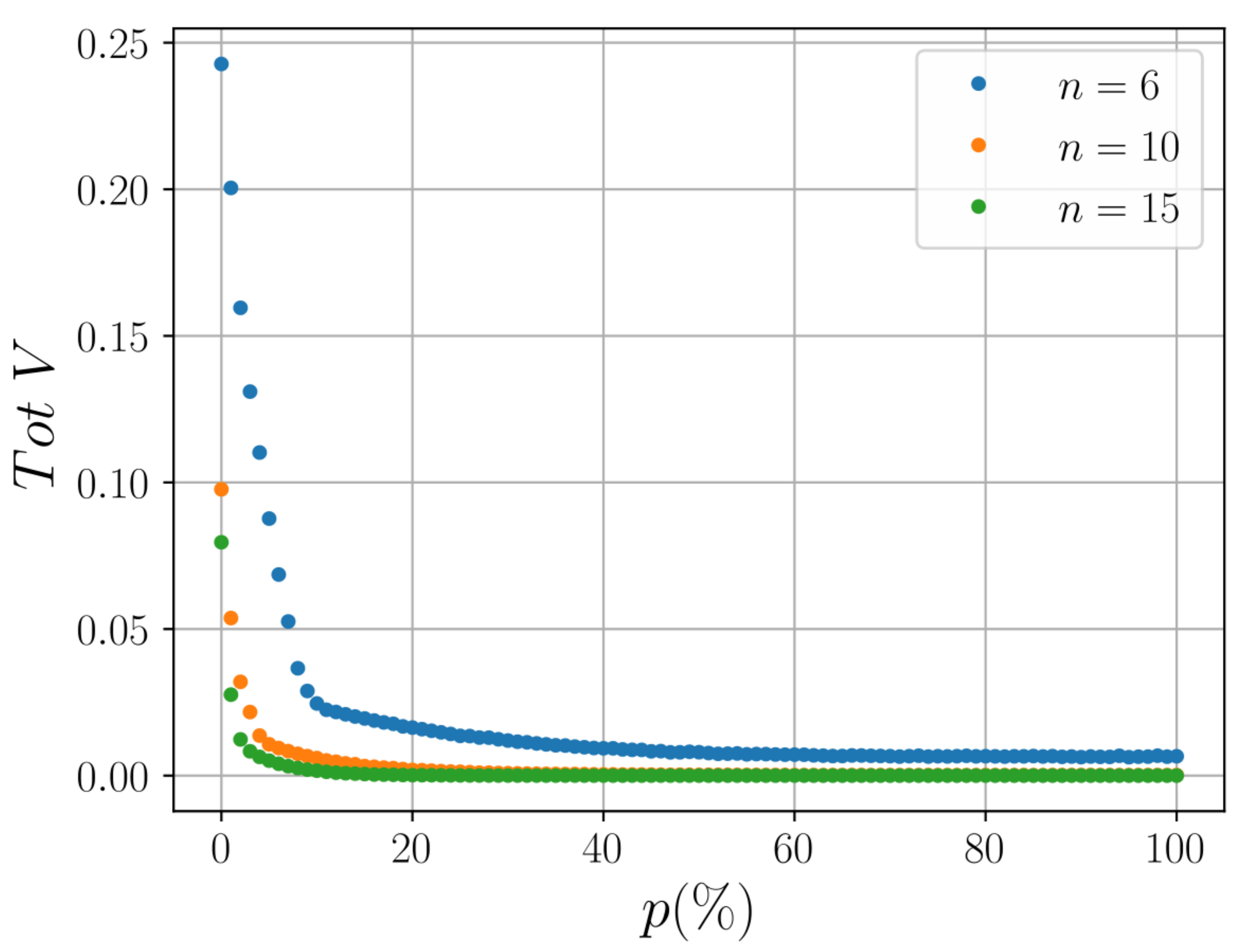}
\caption{\textbf{Numerical simulation of coincidences matrix}. Step 15, step 10 and step 6 plot of the average Total Violation, computed over 10000 disorder configurations, as a function of the disorder level $p$.} \label{fig:totviol}
\end{figure}

We study the average behavior of the violation matrix by averaging over many different evolution realizations with randomness $p$. The average probability distribution is used to obtain the output violation matrices. We show the violation matrix of an indistinguishable photon pair after $15$ steps, for the ordered case (Fig.~\ref{fig:ind_viol}\textbf{a}), corresponding to the disorder level $p=0$, and the completely disordered one (Fig.~\ref{fig:ind_viol}\textbf{b}) featured by maximal randomness $p=1$. In this case, the disorder level $p$ is a relevant quantity, since it indicates the average quantity of disorder imposed on the evolution. As can be seen, violations are present both in the ordered and in the disordered case, though there is an evident migration of the violating values towards the matrix tails.\\
We report in Fig.~\ref{fig:ind_distr} the values of probability $P_{ij}$ of finding a coincidence between two photons emerging from mode $i$ and $j$. By comparing the violation and coincidences matrix, quantum correlations appear between modes with low coincidences and less populated. This consideration nourishes the idea that correlations specifically generated by the QW dynamics are classical and may even smother the underlying non-classical correlations \cite{knight2004propagating,jeong2004simulation}. Therefore, the latter can be detected only where the evolution funnels low population and correlations.\\
Here, the dynamical average behavior of quantum correlation has been studied through Total Violation, defined as the sum of all the $V_{i,j}$ of the violating coincidences. It has been considered as a measure of the total quantum correlation present in the system. The normalized trend in function of the disorder level is reported in Fig.~\ref{fig:totviol}, for different evolution time lengths.
Total Violation has a decreasing trend as $p$ increases, which can be seen as a consequence of the migration picture described above. Since violations are bound to appear only between scarcely populated modes, the global quantum correlations diminishes. Nevertheless, this non-classicality degradation can be challenged by specific single realizations of disorder, as reported in the main text.\\ 
In general, the average evolution appears naturally featured by a decrease of global non-classicality in time.

\section{Numerical results for Single realization}
\begin{figure}[!h] 
\centering
\includegraphics[width=0.45\textwidth]{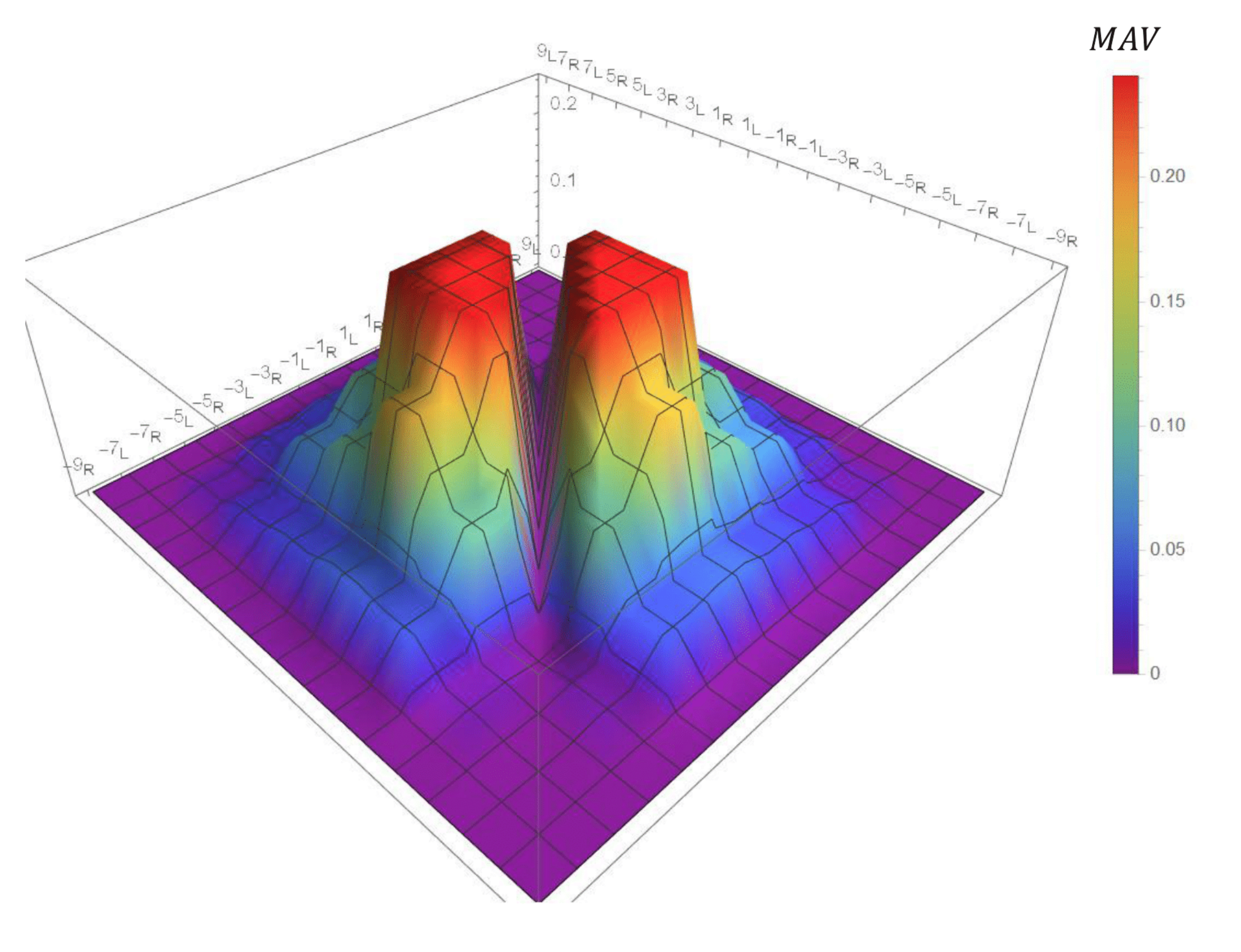}
\caption{\textbf{Numerical simulation of violation landscape}. Plot of he maximum violation achievable for each output modes pair at step 9, obtained by comparing $10^6$ different disorder realizations.} \label{fig:9step_violmat}
\end{figure}

The highest MAV, besides the first step, is achieved at the output of the $9$th step: the MAV for each modes pair $(i,j)$ of the $9$th step output distribution was computed, by analyzing $10^6$ different phase maps each. The resulting landscape in Fig.~\ref{fig:9step_violmat} shows that this maximum can be achieved in different positions, depending on the chosen enhancing disorder configuration. In particular, it suggests that the proper MAV can be achieved only in "central" modes pairs: the MAV can be induced between different mode pairs by imposing different (yet equally enhancing) phase maps, although that is not possible in modes which have not interfered enough. In fact, photons emerging from central modes will have the most interfering paths, getting to be more affected by inhomogeneities along the evolution. This phenomenon is quite understandable by considering the underlying network structure of the evolution (Fig. 1 of the main text). Indeed, central modes are subjected to more complex interference phenomena with respect to the ones close to the boundaries, even more complex when their correlations are considered. This can be directly linked to the amount of MZIs jointly travelled by photons emerging from the two selected modes, i.e. the amount of phase shifts which are imposed over both photons. Therefore, regarding central modes, the manipulation of non-classicality results more powerful and effective.\\ 

\section{Experimental Setup Details}
\begin{figure}[!h] 
\centering
\includegraphics[width=0.425\textwidth]{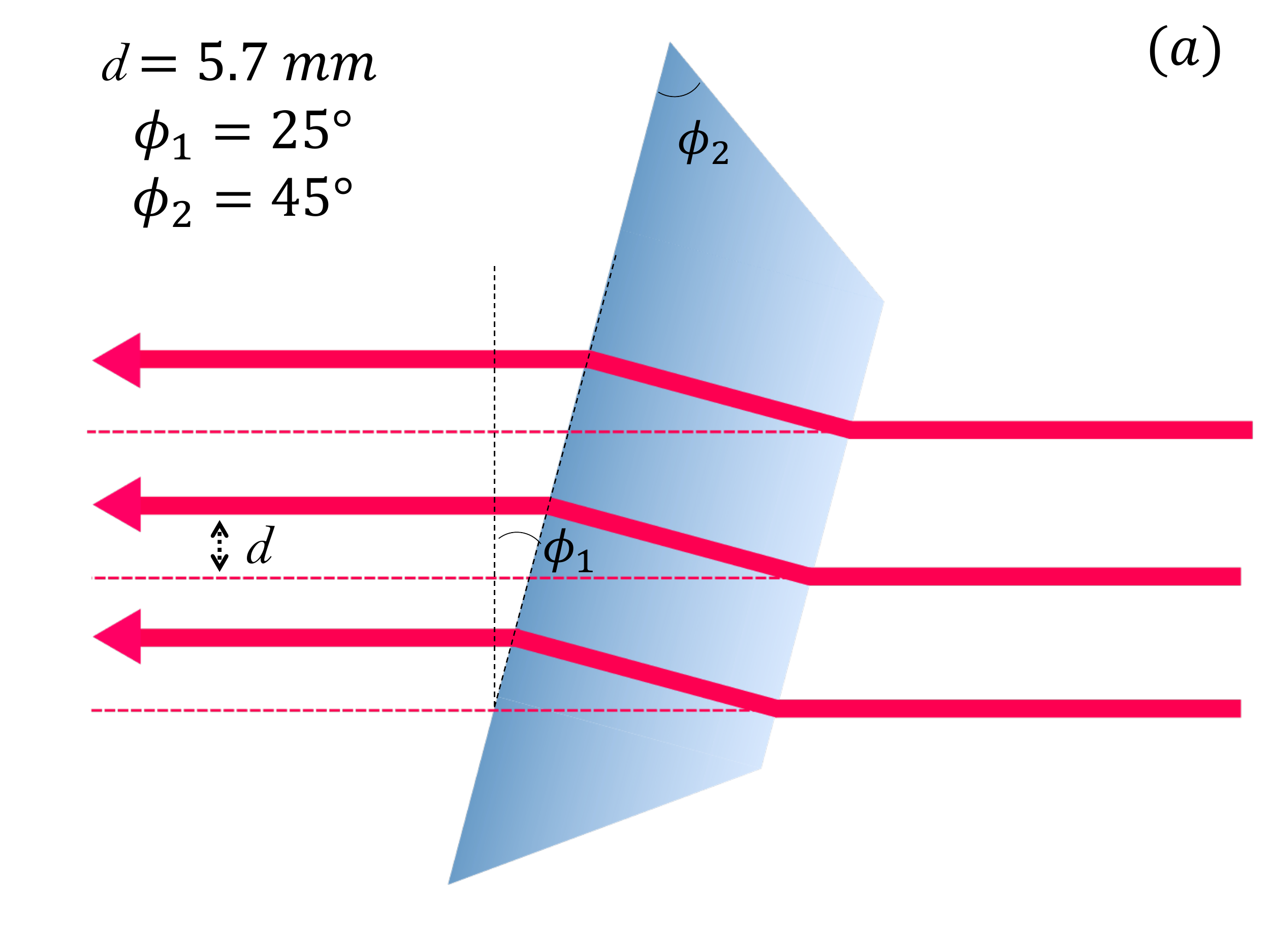}
\includegraphics[width=0.5\textwidth]{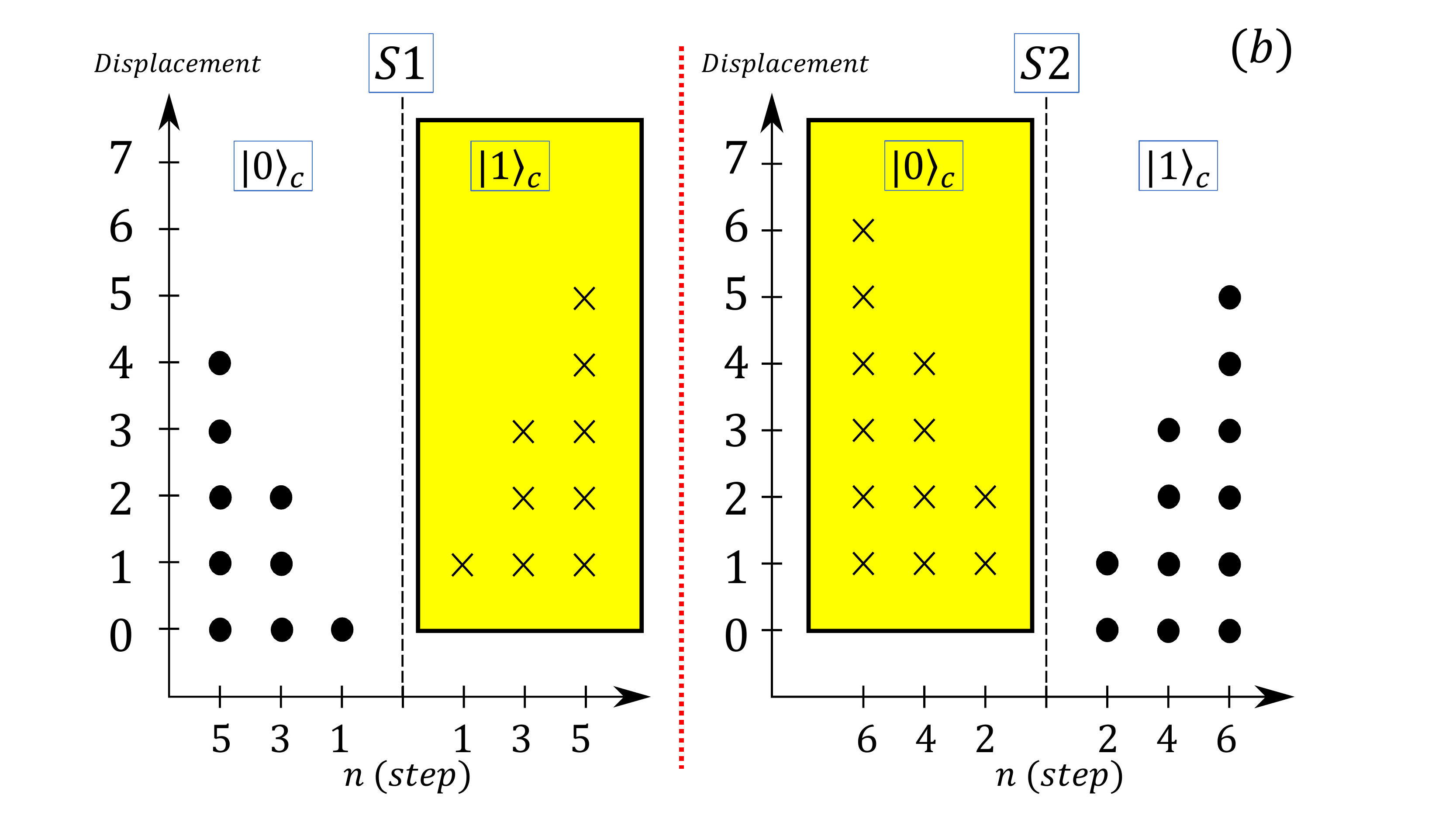}
\caption{\textbf{Beam displacer and spatial structure of the QW}. \textbf{a)} A sketch of the Beam Displacer (BD) functioning, provided with relevant geometrical parameters. Through this device, it is possible to realize \textbf{b)} a spatial structure featured by a distinct localization of each QW mode at any evolution step in the plane orthogonal to the propagation direction.}
\label{fig:displacement}
\end{figure}

The exploitation of the \textit{z}-axis for the realization of the Quantum Walk network relies on Beam Displacers (Fig. \ref{fig:displacement}\textbf{a}), implementing the unique spatial distribution of the network nodes (Fig. \ref{fig:displacement}\textbf{b}), essential to the actual realization of a space-time disorder.

Photon pairs are generated by a source realized according to the model described in \cite{fedrizzi2007wavelength}: a PPKTP crystal, embedded in a Sagnac interferometer, pumped by a CW laser radiation ($\lambda_p = 405$nm), which generates collinear pair of photons (idler and signal) with opposite polarization at a wavelength $\lambda_{i,s} = 2\lambda_p = 810$nm . They are coupled to a pair of optical fibers and experience an additional path through air before starting the actual QW evolution.
They are made indistinguishable in all possible degrees of freedom such as polarization, wavelength, and propagation mode. In particular, we tuned the relative free-space path to get the best bunching effect when they impinge on the bulk-optics BS for the first time, corresponding to the first step of the QW. The unavoidable, critical free-space adjustement of beam superposition on the BS mainly limits the achieved visibility of HOM effect to an average $V\sim 89\%$.\\

\section{Experimental Results} 

\begin{figure*}[t] 
\centering
\includegraphics[width=1\textwidth]{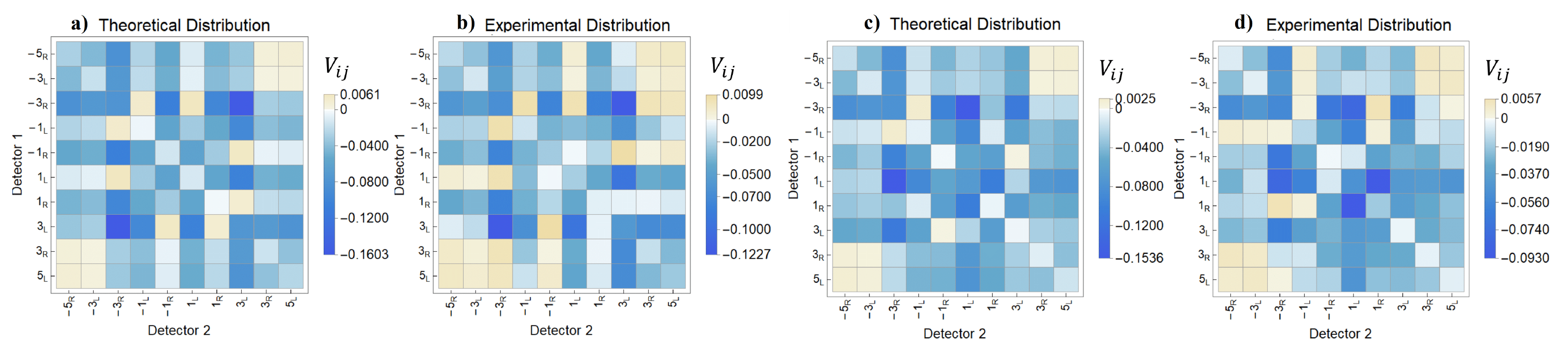}
\caption{\textbf{Violation matrices at the 5th step.} Theoretical and experimental violation matrices at the 5th step for \textbf{a) and b)} ordered evolution and \textbf{c) and d)} disordered evolution. The disorder configuration has been chosen randomly.} \label{fig:5th_step}
\end{figure*}

\begin{figure*}[!t] 
\centering
\includegraphics[width=1\textwidth]{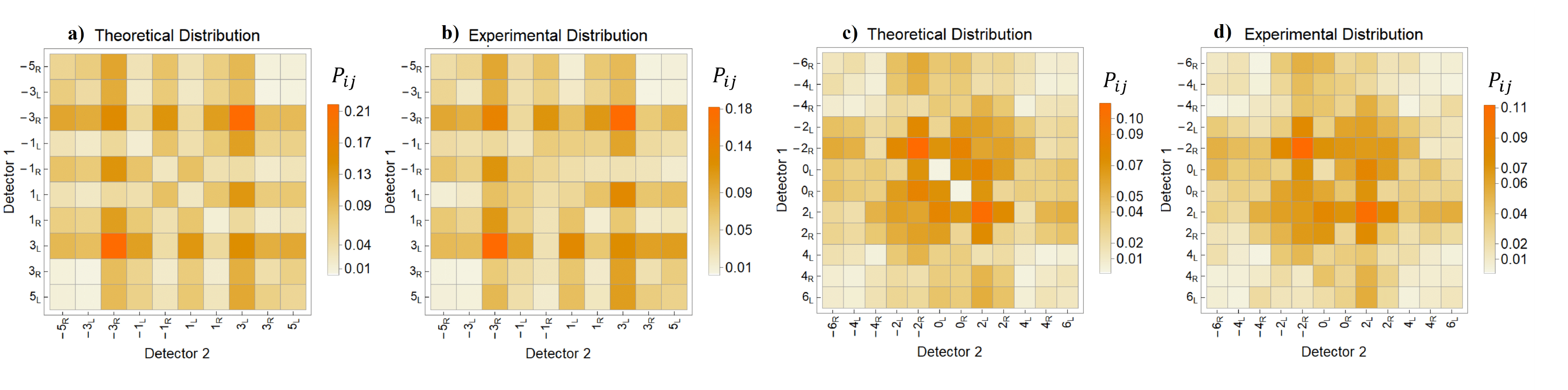}
\caption{\textbf{Numerical and experimental coincidences matrices}. $5$th step \textbf{a)}  theoretical coincidences matrix and \textbf{b)} corresponding experimental measurement output for the ordered quantum walk.  6th step \textbf{c)}  theoretical coincidences matrix and \textbf{d)} corrsponding experimental measurement output for the optimal disordered configuration.  The similarities between theoretical and experimental coincidences distributions are $98(\pm1)\%$ for the 5th step case (\textbf{a-b}) and $97.5(\pm1.3)\%$ for the 6th step case (\textbf{c-d}). Errors are computed accounting for Poissonian statistics of measured coincidences.} \label{fig:coinc_matrices}
\end{figure*}

The output violation distributions for the $5$th step are shown in Fig.~\ref{fig:5th_step}. They provide a preliminary demonstration of the dependence of quantum correlation distributions from the disorder pattern imposed on the evolution. The experimental results appears to be not in perfect agreement with the expected ones, since, in this case, the error on $V_{i,j}$ values is large with respect to the measured absolute values of violation. 
It is useful to observe the corresponding coincidence matrices for the ordered $5$th step and the optimal $6$th step (Fig.~\ref{fig:coinc_matrices}). Disorder, as demonstrated in several previous works \cite{schreiberecoherence2011,crespi2013anderson,fedrizzi2007wavelength,geraldi2019experimental}, determines first an effect of spread hindering, also for multiparticle systems. This effect can be noticed even in the case of a single disorder configuration (Fig.~\ref{fig:coinc_matrices}). The manipulation of this localization effect can change the non-classicality pattern in the output distribution in many different configurations, changing the probability of finding coincident photons between the output modes. Indeed, the corresponding experimental coincidences distributions result in good agreement with the expected ones .

\end{document}